\documentclass[reprint,aps,prb,showpacs]{revtex4-1}
\usepackage{color,graphicx,calc,url}
\usepackage{dcolumn}
\usepackage{bm}
\usepackage{amssymb}
\usepackage{amsmath}
\usepackage{wasysym}
\usepackage{natbib}
\usepackage{mathrsfs}

\begin{document}
\raggedbottom
\title{Thermal conductivity of thin insulating films determined by tunnel magneto-Seebeck effect measurements and finite-element modeling}
\author{Torsten Huebner,$^1$ Ulrike Martens,$^2$ Jakob Walowski,$^2$ Markus M\"unzenberg,$^2$ Andy Thomas,$^3$ G\"unter Reiss,$^1$ Timo Kuschel,$^{1}$
\email{Electronic mail: thuebner@physik.uni-bielefeld.de}}
\affiliation{$^1$Center for Spinelectronic Materials and Devices, Department of Physics, Bielefeld University, Universit\"atsstra\ss e 25, 33615 Bielefeld, Germany\\
$^2$Institut f\"ur Physik, Greifswald University, Felix-Hausdorff-Strasse 6, 17489 Greifswald, Germany\\
$^3$Leibniz Institute for Solid State and Materials Research Dresden (IFW Dresden), Institute for Metallic Materials, Helmholtzstrasse 20, 01069 Dresden, Germany}

\date{\today}

\keywords{}

\begin{abstract}

In general, it is difficult to access the thermal conductivity of thin insulating films experimentally just by electrical means. Here, we present a new approach utilizing the tunnel magneto-Seebeck effect (TMS) in combination with finite-element modeling (FEM). We detect the laser-induced TMS and the absolute thermovoltage of laser-heated magnetic tunnel junctions with 2.6\,nm thin barriers of MgAl$_2$O$_4$ (MAO) and MgO, respectively. A second measurement of the absolute thermovoltage after a dielectric breakdown of the barrier grants insight into the remaining thermovoltage of the stack. Thus, the pure TMS without any parasitic Nernst contributions from the leads can be identified. In combination with FEM via COMSOL, we are able to extract values for the thermal conductivity of MAO ($0.7$\,W/(K$\cdot$m)) and MgO ($5.8$\,W/(K$\cdot$m)), which are in very good agreement with theoretical predictions. Our method provides a new promising way to extract the experimentally challenging parameter of the thermal conductivity of thin insulating films. 

\end{abstract}

\maketitle

Within the upcoming stage of spin caloritronics [\onlinecite{Bauer}], the improvement of existing memory devices, for example via waste heat recovery generating thermopower in magnetic tunnel junctions (MTJs), play an important role. After the theoretical description of the basic processes by Czerner et al. [\onlinecite{czerner}], two different experimental approaches were realized to measure the tunnel magneto-Seebeck (TMS) effect. This effect describes the thermovoltage of an MTJ, caused by an applied temperature difference, depending on its' magnetic state. In general, the two experiments differ with regard to the creation of the temperature gradient. On the one hand, Liebing et al. [\onlinecite{Liebing1, Liebing2, Liebing3, Liebing4}] and B\"ohnert et al. [\onlinecite{drop2, boehnert2}] used a patterned, electrically isolated heater line on top of the MTJs to generate a temperature difference across the layer stack. Accordingly, this effect is called the extrinsic heating TMS effect. On the other hand, Walter et al. [\onlinecite{walter}] utilized a laser beam to create the temperature difference, which is referred to as the laser-induced TMS effect.

In the course of subsequent laser-induced TMS studies, insulating substrates (such as MgO) were identified to exhibit no parasitic capacitive effects in contrast to semiconducting substrates such as Si [\onlinecite{Boehnke1}], an additional bias voltage resulted in giant effect ratios of several thousand percent [\onlinecite{Boehnke2}] and Heusler electrodes showed large thermovoltages and enhanced switching ratios in comparison to commonly used electrodes such as CoFeB [\onlinecite{Boehnke3}]. Additionally, the laser-position dependence of the TMS effect was investigated with a focused laser beam [\onlinecite{TMS}], a direct comparison of the laser-induced and a proposed intrinsic TMS effect, was conducted [\onlinecite{huebner}] and the influence of the barrier material (MgAl$_2$O$_4$ (MAO) vs. MgO) and its thickness was studied [\onlinecite{huebner2}]. In the latter publication, MAO as a barrier material was found to exhibit larger thermovoltages in comparison to MgO. For both MAO and MgO MTJs, a maximum TMS ratio has been found using a barrier thickness of 2.6\,nm.

All studies of the TMS effect have a major drawback, which is the lacking knowledge of the real temperature distribution within the MTJs. Usually, this distribution and the resulting temperature difference, which is needed for the calculation of absolute Seebeck coefficients of the MTJs, is approximated by, e.g., finite-element modeling (FEM) via COMSOL (COMSOL Multiphysics Reference Manual, version 4.4, COMSOL, Inc, www.comsol.com). First attempts to measure the temperature on top and at the bottom of the layer stack were performed, but are still relying on additional COMSOL simulations [\onlinecite{boehnert2}]. 

A critical parameter of these simulations is the thermal conductivity of the thin barriers. In case of insulating films, this quantity is not directly accessible via, for example, the 3$\omega$ method [\onlinecite{cahill}] without any additional sophisticated data processing [\onlinecite{euler}]. Recently, Cahill and Kimling [\onlinecite{kimling2}]  developed a thermal reflectance method to experimentally determine the thermal conductivity of thin insulating films based on the work shown in Ref. [\onlinecite{cahill3}]. Preliminary results for thin MgO barriers are in good agreement with theoretical works (see Tab. \ref{thermal_cond}). 

In general, the thermal conductivity of thin films is reduced by about an order of magnitude in comparison to bulk values due to the reduced number of available phonon states [\onlinecite{cahill2}]. However, Zhang et al. [\onlinecite{drop}] proposed that the actual thermal conductivity of ultra thin films might be an additional order of magnitude below the previously assumed values due to an imbalance of the electron and phonon temperature at magnetic interfaces on the nano scale, respectively. Table 1 summarizes the experimental and theoretical values of the thermal conductivities of MAO and MgO (which we choose as barrier materials) of both the bulk and thin film regime. Due to the lack of measurements in case of thin MAO films and the discrepancies between the measurements and the theoretical predictions in case of MgO, we assume a value of $0.2-2.3$\,W/(K$\cdot$m) for MAO and $0.4-4$\,W/(K$\cdot$m) for MgO as a basis for our FEM.

\begin{figure}[bt]\centering
		\includegraphics{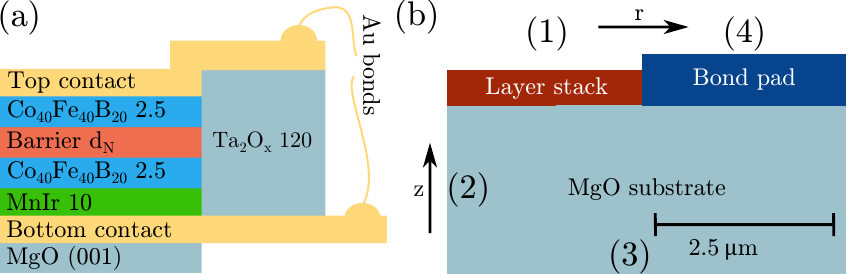}
	\caption{(a) MTJ structure with top and bottom electrode consisting of Co$_{40}$Fe$_{40}$B$_{20}$. Additionally, the bottom electrode is pinned by Mn$_{83}$Ir$_{17}$. Numbers are thicknesses in nanometers. (b) Schematic of the used COMSOL model, all layers are individually incorporated in the red layer stack. The laser, i.e. the heat source, is applied to the layer stack (1). To reduce computational time, the model is rotationally symmetric around the z-axis (2). The bottom of the substrate is kept at a constant temperature of 293.15\,K (3), while the remaining boundaries are assumed to be thermally insulating (4).}
	\label{fig:sim}
\end{figure}

The basic structure of the MTJs is schematically shown in Fig. \ref{fig:sim}(a). They are deposited with a varying barrier thickness of MgO and MAO between 1\,nm and 3\,nm and prepared as described in Ref. [\onlinecite{huebner2}]. Two layers of Co$_{40}$Fe$_{40}$B$_{20}$ are used as ferromagnetic electrodes, while Mn$_{83}$Ir$_{17}$ ensures a higher switching field of the bottom electrode via the exchange bias effect. In addition, we assume a thermal conductivity of $6$\,W/(K$\cdot$m) of the MnIr layer, which is based on the Wiedemann-Franz law [\onlinecite{mnir}].

Figure \ref{fig:sim}(b) depicts the used COMSOL model, including the layer stack (top/bottom contacts, electrodes and barrier) (red), the Au bond pad (blue) and the MgO substrate. Furthermore, the heat source (1), the rotationally symmetric z-axis (2), the constant temperature of 293.15\,K of the bottom of the substrate (3) and the thermally insulating boundaries (4) are shown. 

 \begin{table}[bt]\centering
	\caption{Experimental and theoretical thermal conductivities of MAO and MgO bulk and thin film samples at room temperature with the resulting range of the thermal conductivities assumed in this work.}
	\label{thermal_cond}
		\begin{tabular}{l| c c}
		\hline\hline
			 & $\kappa$[MAO] (W/(K$\cdot$m)) & $\kappa$[MgO] (W/(K$\cdot$m)) \\
			\hline
			Ref. \onlinecite{MAO_bulk_1, MAO_bulk_2,MgO_bulk} (bulk)& $22-24$ & $48$\\
			Ref. \onlinecite{MgO_thin} (exp.)& --- & $4$\\
			Ref. \onlinecite{kimling2} (exp.)& --- & 0.5\\
			Ref. \onlinecite{drop} (theo.)& --- & $0.4$\\
			This work & $0.2-2.3$ & $0.4-4$\\
			\hline \hline
		\end{tabular}
\end{table}

\begin{figure}[bt]\centering
		\includegraphics[width=\columnwidth]{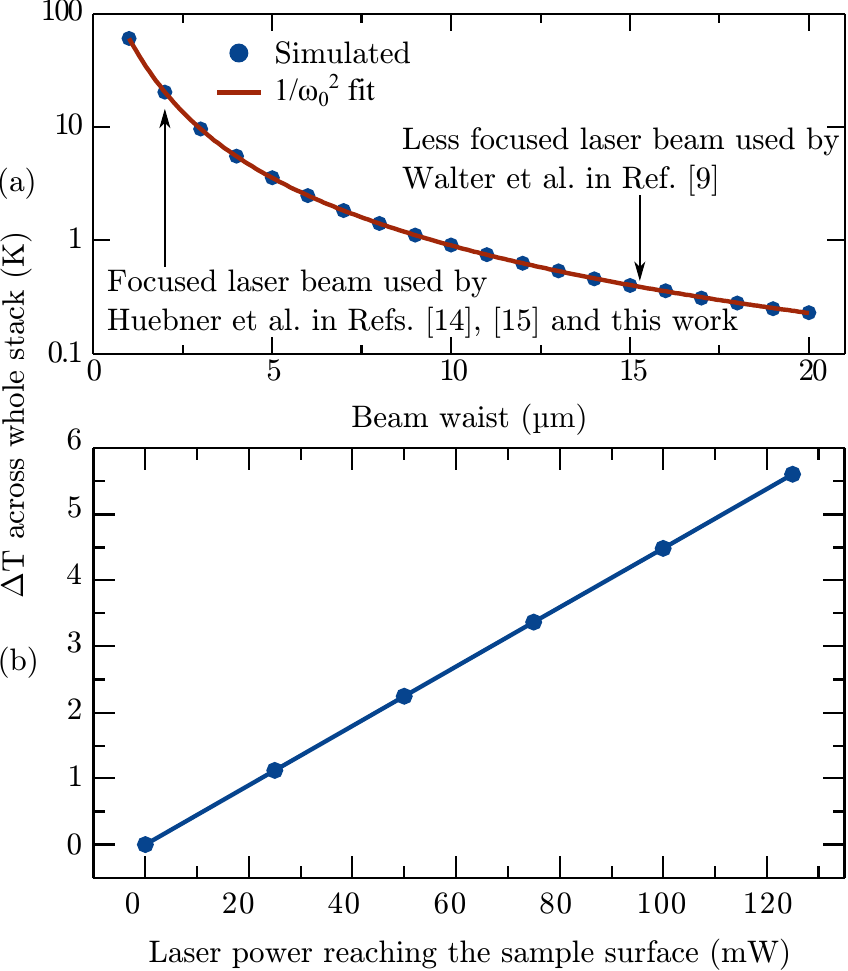}
	\caption{Exemplary results of the FEM for an MTJ with a 1.8\,nm MAO barrier. (a) Influence of the beam waist on the resulting temperature difference across the whole stack with a maximum laser power of 124\,mW and a $1/\omega_0^2$ best fit. The arrows mark the values used in Refs. [\onlinecite{huebner,huebner2}] and this work $(\omega_0=(1.92\pm0.01)\,\mu$m) and in the initial laser-TMS publication [\onlinecite{walter}] ($\omega_0\approx 15\,\mu$m), respectively. (b) Resulting temperature differences with the focused laser beam ($(\omega_0=(1.92\pm0.01)\,\mu$m)) across the whole stack for different laser powers. Here, the actual laser power reaching the sample surface is used for the model.}
	\label{fig:comsol1}
\end{figure}

In order to verify the validity of our model, Fig. \ref{fig:comsol1}(a) visualizes the resulting temperature differences across the whole stack depending on the beam waist of the laser. Here, the obtained temperature difference of 50\,mK for a beam waist of around 15\,$\mu$m used by Walter et al. [\onlinecite{walter}] is reproduced. In our case, the laser is focused down to a minimum beam waist of around 2\,$\mu$m, which results in relatively large temperature differences in the range of 10\,K when heating MTJs with a similar size of $0.5\,\pi\mu$m$^2$. Additionally, Fig. \ref{fig:comsol1}(b) shows the linear dependence of the temperature difference when linearly increasing the laser power, as expected from experiments [\onlinecite{huebner, TMS}]. The laser power used within the simulations is extracted from power calibration measurements directly above the sample surface.

Figures 3(a,b) depict the thermal profiles across the layer stack for the lower (0.3\,W/Km) and upper (2.3\,W/Km) limit of the thermal conductivity of MAO, respectively. Here, the stack position of 0\,nm corresponds to the top of the stack. Furthermore, and in accordance with the results presented in Ref. [\onlinecite{huebner2}], the temperature distributions for barrier thicknesses ranging from $(1.4 \text{ to } 2.6)$\,nm are shown. The temperature of the heated electrode rises linearly with increasing barrier thickness, since the barrier acts as a thermal resistance. In addition, the temperature difference across the barrier ($\Delta$T MAO) rises linearly as well. 

\begin{figure}[bt]\centering
		\includegraphics[width=\columnwidth]{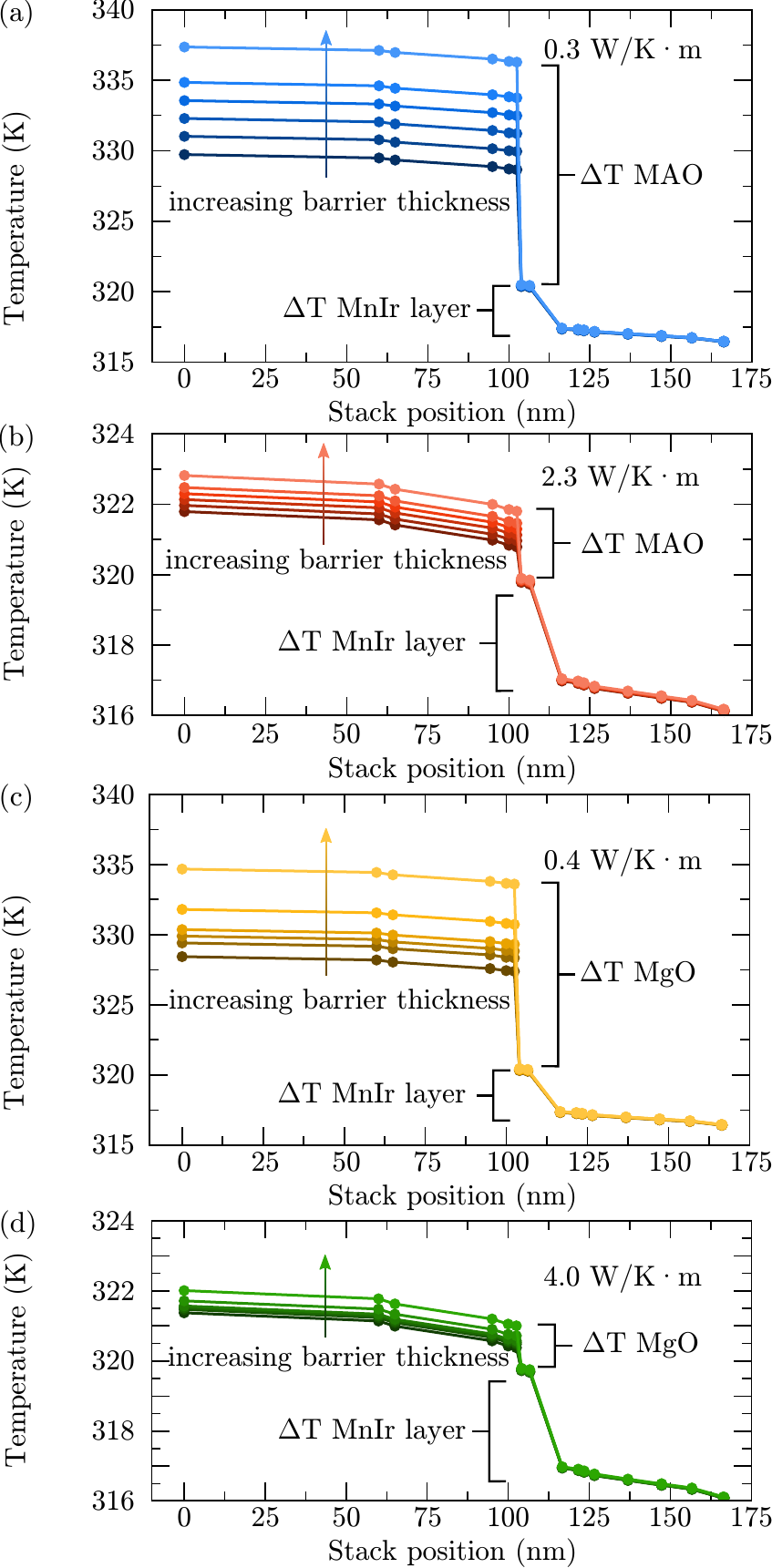}
	\caption{FEM results for MTJs with MAO (MgO) barrier thicknesses ranging from $1.4\,$nm to $2.6$\,nm ($1.6\,$nm to $2.9$\,nm). Thermal profiles across the layer stack with thermal conductivities of the MAO barrier of (a) 0.3\,W/(K$\cdot$m) and (b) 2.3\,W/(K$\cdot$m) and the MgO barrier with thermal conductivities of (c) 0.4\,W/(K$\cdot$m) and (d) 4\,W/(K$\cdot$m).}
	\label{fig:comsol2}
\end{figure}

\begin{figure}[bt]\centering
		\includegraphics[width=\columnwidth]{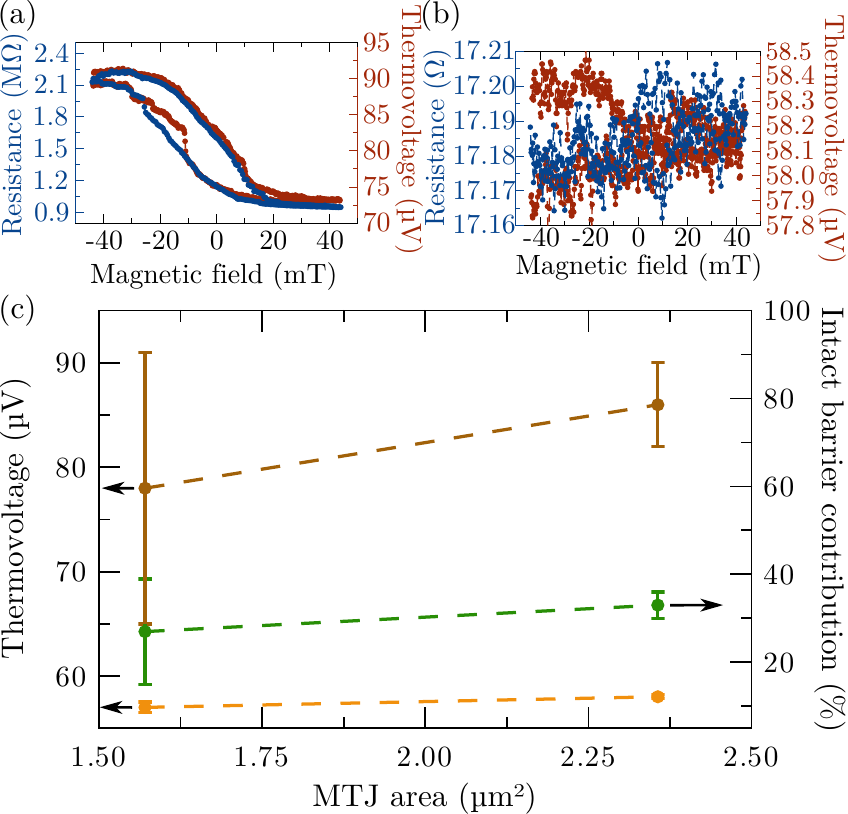}
	\caption{(a) TMR and TMS minor loops of an MTJ with a 2.6\,nm MgO barrier. The TMR (TMS) effect amounts to 126\,\% (23\,\%). (b) Remaining resistance and thermovoltage after a voltage of 3\,V is applied to the junction. (c) Absolute thermovoltages (dark orange) and remaining thermovoltages after the dielectric breakdown of the junction (light orange) depending on the junction size (left scale) and contribution of the intact MgO barrier to the absolute thermovoltage (green, right scale). The errors result from averaging over several MTJs.}
	\label{fig:breakdown_final}
\end{figure}

Besides the temperature difference across the barrier, another significant temperature difference builds up at the MnIr layer. These two temperature differences in the range of a few K generate most of the absolute thermovoltage, since the remaining temperature differences are negligible in comparison. With the measurements of the absolute thermovoltage before and after a dielectric breakdown of the barrier as shown in Ref. [\onlinecite{huebner2}] in case of MAO (MgO data shown in Fig. \ref{fig:breakdown_final}(c)), we are able to relate the simulated temperature differences of the MAO/MgO and the MnIr layer to the intact barrier (MAO/MgO and MnIr temperature difference contributing, namely $V_{\text{MAO,MgO}}$) and the destroyed barrier (only MnIr temperature difference contributing, namely $V_{\text{MnIr}}$). Thus, we are able to deduce the thermal conductivity of the barrier via the ratio of $V_{\text{MAO,MgO}}$ and $V_{\text{MnIr}}$. 

Figure \ref{fig:breakdown_final} shows the measurements of the TMR and TMS effect with an intact barrier of 2.6\,nm MgO (see Fig. \ref{fig:breakdown_final}(a)) and the destroyed barrier, after a voltage of 3\,V is applied to the junction (see Fig. \ref{fig:breakdown_final}(b)). The working MTJ shows a TMR (TMS) effect of 126\,\% (23\,\%). After the dielectric breakdown, the signal is no longer depending on the external magnetic field, while the resistance changes from the M$\Omega$-regime to the $\Omega$-regime. In addition, Fig. \ref{fig:breakdown_final}(c) shows the remaining thermovoltage depending on the MTJ area.

In the regime of homogeneously heating the MTJs (cf. Refs. [\onlinecite{TMS, huebner2}]), the contribution of the intact barrier amounts to 70\,\% in case of an MAO barrier ($V_{\text{MAO}}$=$\frac{7}{3}V_{\text{MnIr}}$, cf. Ref. [\onlinecite{huebner2}], Fig. 4(a) therein) and to about 30\,\% in case of an MgO barrier ($V_{\text{MgO}}$=$\frac{11}{39}V_{\text{MnIr}}$, cf. Figs. \ref{fig:breakdown_final}(b,c)). Hence, we can deduce the thermal conductivity of both barrier materials via the assumptions of the temperature difference at the MnIr layer and the thermovoltage scaling inverse proportionally with the thermal conductivity. Taking the thicknesses of the layers into account results in a thermal conductivity of MAO of 0.7\,W/(K$\cdot$m), which is well within the limits assumed for our simulations. However, the same procedure yields a thermal conductivity of MgO of 5.8\,W/(K$\cdot$m), which is even above the value of 4\,W/(K$\cdot$m) experimentally determined by Lee et al. [\onlinecite{MgO_thin}] for crystallites with a size of 3\,nm to 7\,nm.

In conclusion, we have shown that a combination of laser-induced TMS measurements and COMSOL simulations of the resulting thermal profiles offer a new approach to access thermal conductivities of thin insulating films. Additionally, this approach has the potential to establish a new method to access information regarding the thermal distribution inside nanometer thin layer stacks. For example, if a certain temperature distribution and the resulting thermovoltage of a material system is known, it might serve as a standard for an additional layer, which is not known in terms of its' thermal conductivity. Furthermore, this approach stresses the importance of measurements of the thermovoltage with both intact and electrically destroyed barrier. However, our method does not include thermal interface resistances, which can play a vital role in the thin film regime. The influence of this neglect has to be taken into account in future simulations and experiments.

\section{Acknowledgments}

The authors gratefully acknowledge financial support from the Deutsche Forschungsgemeinschaft (DFG) within the priority program Spin Caloric Transport (SPP 1538, MU 1780/8-2, KU 3271/1-1, RE 1052/24-2, TH 1399/4-2).

\end{document}